\documentclass[twocolumn,apl,amsmath,amssymb,showpacs]{revtex4-2}
\usepackage{epsf}      
\usepackage{graphicx}
\usepackage{color}
\usepackage{soul}
\usepackage{gensymb}
\usepackage{sidecap}
\usepackage{amsmath}
\usepackage{mathtools}
\usepackage[hidelinks,colorlinks=true,linkcolor=blue,citecolor=blue]{hyperref}

\usepackage[hidelinks,colorlinks=true,linkcolor=blue,citecolor=blue]{hyperref}
\DeclareUnicodeCharacter{2212}{\textendash}

\begin{document}
\title{Revealing the interfacial kinetic mechanisms in high-entropy doped Na$_3$V$_2$(PO$_4$)$_3$ through electrochemical investigation and distribution of relaxation times}
\author{Manish Kr.~Singh}
\affiliation{Department of Physics, Indian Institute of Technology Delhi, Hauz Khas, New Delhi-110016, India}
\author{Rajendra S.~Dhaka}
\email{rsdhaka@physics.iitd.ac.in}
\affiliation{Department of Physics, Indian Institute of Technology Delhi, Hauz Khas, New Delhi-110016, India}

\date{\today} 

\begin{abstract}

We designed a high-entropy doped NASICON cathode, Na$_3$V$_{1.9}$(CrMoAlZrNi)$_{0.1}$(PO$_4$)$_3$ and investigate its electrochemical performance for sodium-ion batteries (SIBs) to understand the diffusion mechanism including distribution of relaxation times analysis of interfacial kinetics. This trace doping induces high-entropy mixing at the vanadium site, tuning the lattice and enhancing specific capacity, activating V$^{4+}$/V$^{5+}$ redox couple 3.95~V. Interestingly, it delivers a reversible capacity of 119~mAh~g$^{-1}$ at 0.1~C, and demonstrate excellent stability of 68\% after 1000 cycles at 10~C. The calculated diffusion coefficient values are found within the range of \(10^{-11}\)--\(10^{-13}~\mathrm{cm^2\,s^{-1}}\). The systematic investigation of temperature and voltage-dependent impedance data using the distribution of relaxation times provides deeper insights into the underlying charge-transfer and transport processes. The full cells with hard carbon delivers 326~Wh~kg$^{-1}$ (with respect to cathode mass) at $\approx$3.2~V and retained $\sim$79\% capacity after 100 cycles at 2~C. Our study opens new avenues for developing high-entropy doped cathodes for enhanced structural stability, extended redox activity, and optimized electrochemical kinetics for practical implementation of SIBs. 
\end{abstract}

\maketitle
\section{\noindent Introduction}

The rising demand for sustainable energy storage solutions has led to extensive research in next generation battery technologies beyond lithium-ion batteries (LIBs) \cite{Hirsh_AEM_20, Delmas_AEM_18}. In this regard, sodium-ion batteries (SIBs) have emerged as a promising candidate owing to the natural abundance and low cost resources of sodium precursors \cite{Hirsh_AEM_20}. However, the practical deployment of SIBs remains hindered by challenges such as limited energy density, sluggish ion diffusion, and structural instability \cite{Hwang_CSR_17} where cathode materials play a crucial role in determining the electrochemical performance. In recent years, polyanionic cathodes (e.g., phosphates, sulfates, and pyrophosphates) have gained significant attention for their stable crystal structures, high thermal stability, and voltage tenability due to strong covalent bonding \cite{Xu_ACSCS_23_23}, but they exhibit lower electronic conductivity compared to layered oxides \cite{Jian_AEM_13}. Among the polyanionics, NASICON (Na Super Ionic CONductor)-type materials, particularly Na$_3$V$_2$(PO$_4$)$_3$ (NVP), have been considered as potential cathodes for SIBs due to their three-dimensional (3D) open framework, excellent ionic conductivity, and high thermal stability. The NVP cathode possess two redox reactions at 3.4~V and 3.9~V vs.\ Na/Na$^+$, corresponding to the V$^{3+}$/V$^{4+}$ and V$^{4+}$/V$^{5+}$ redox transition, respectively \cite{Jian_AEM_13}. The latter being irreversible due to structural instability \cite{Jian_AEM_13}, while still offering a theoretical capacity of approximately 115 mAh g$^{-1}$ and an energy density of around 390 Wh kg$^{-1}$ using V$^{3+}$/V$^{4+}$ redox couple. However, their intrinsic electronic conductivity is poor due to insulating nature of PO$_{4}$$^{3-}$ unit, affecting the electronic de-localization of –V–O–V– \cite{Sapra_ESM_2025}. Consequently, various strategies have been explored to enhance its electrochemical performance, including cation doping, carbon coating, tuning particle size, etc. \cite{Jiang_AEM_15, Shen_AMI_2016, Zhang_AMI_2019, Jiang_JMCA_2016}. For instance, electrochemically inactive ions such as Al$^{3+}$, Ni$^{2+}$, and Zr$^{4+}$ act as structural pillars, suppressing volume expansion and strengthening the polyanionic framework through enhanced M--O bonding and lattice modulation \cite{Sun_EEM_22, Li_AMI_16, Guo_JEC_2022}. Also, transition metals like Cr$^{3+}$ and Mo$^{6+}$ serve to tune the electronic environment to facilitates redox activation at higher potentials \cite{Chen_JACS_21} or induce sodium vacancies that significantly accelerate ion diffusion kinetics \cite{Li_JMCA_2018}. 

However, in this context, high-entropy (HE) doping engineering has emerged as an innovative strategy where multiple transition metal cations are introduced into the cathode framework \cite{Garcia_ESM_24,  Ma_EES_21, Zhou_EES_25, Su_AEM_2025, Gu_AM_2022, Wang_JMCA_2024, Zhou_Nano_En_24}. This approach enhances the configurational entropy of the material, suppresses the phase separation, and stabilizes multi-electron redox reactions \cite{Zhang_AFM_25, Hou_ESM_24, Sun_ACS_nano_25, Li_AS_22}. The HE doped NASICON cathodes are expected to exhibit improved electronic conductivity, enhanced Na$^+$ migration kinetics, and excellent cycling stability even under deep de-sodiation conditions \cite{ Hou_ESM_24, Li_AFM_24, Zhang_AFM_25, Hou_ESM_24, Zhang_ESM_24}. Recently, Du \textit{et al.} synthesized Na$_{3.45}$V$_{0.4}$(Fe$_{0.4}$Ti$_{0.4}$Mn$_{0.45}$Cr$_{0.35}$)(PO$_4$)$_3$ via ultrafast high-temperature shock, ensuring phase stability and reversible multi-electron transfer (2.4/2.8 e$^{-}$), which delivers 137.2/162.0 mAh g$^{-1}$ (at 0.2 C) within 2.0--4.45/1.5--4.45 V (vs. Na$^+$/Na), retains 83.7\% capacity after 2000 cycles at 5 C, and operates from --50\degree C to 60\degree C \cite{Du_AM_25}. Furthermore, synthesis of Na$_{3.4}$(Fe$_{0.2}$Mn$_{0.2}$V$_{0.2}$Cr$_{0.2}$Ti$_{0.2}$)$_2$(PO$_4$)$_2$$_{.98}$F$_{0.02}$ via a sol-gel method, demonstrated 138.8 mAh g$^{-1}$ at 0.5 C, retaining 74\% capacity after 50 cycles \cite{Wang_ICF_24}. Also, the Na$_3$V$_{1.45}$(Fe,Al,Cr,Mn,Ni)$_{0.5}$Mo$_{0.02}$Zr$_{0.03}$(PO$_4$)$_3$, synthesized via a solid-state method, exhibits 93.5\% capacity retention after 100 cycles at 1 C and 80.1\% retention after 800 cycles at 5 C \cite{Liao_ESM_25}. Ding {\it et al.} employed a trace-level incorporation of few dopants, strategically chosen for their distinct chemical characteristics to create a synergistic cocktail effect within the host structure, which activated V$^{4+}$/V$^{5+}$ redox couple at around 4.0 V and the optimized composition provided discharge capacity of 127 mAh g$^{-1}$ with 97\% retaintation after 5000 cycles at 40 C \cite{Ding_ACS_Nano_24}. However, different combinations are yet to be explored with detailed analysis of electrochemical performance to understand diffusion kinetics and interface effects \cite{Zhou_EES_25, Hou_ESM_24}. 

Therefore, in this paper, we use different electrochemical tools and distribution of relaxation times to understand the interfacial kinetics of newly designed Na$_3$V$_{1.9}$(Cr$_{0.02}$Mo$_{0.02}$Al$_{0.02}$Zr$_{0.02}$Ni$_{0.02}$)(PO$_4$)$_3$ (named as NVP-HE) cathode for SIBs. Interestingly, we find that this trace HE doping activated V$^{4+}$/V$^{5+}$ redox couple around 4.0~V and the electrode exhibits a high initial capacity of 119~mAh~g$^{-1}$ at 0.1~C, with an impressively low polarization ($\sim$0.05~V). Moreover, it demonstrates superior rate capability across a wide range of C-rates (0.1~C to 5~C) and notable long-term cycling stability, delivering 93\% capacity retention at 2~C over 100 cycles and maintaining 56~mAh~g$^{-1}$ after 1000 cycles at 10~C. Additionally, sodium-ion transport kinetics evaluated by cyclic voltammetry (CV), galvanostatic intermittent titration (GITT), and electrochemical impedance spectroscopy (EIS) indicates favorable Na$^+$ mobility with diffusion coefficients of 10$^{-11}$--10$^{-13}$~cm$^2$/s. Complementary {\it in-situ} impedance measurements across voltage and temperature, analyzed through distribution of relaxation times, further elucidated charge-transfer and ion-transport processes. The full cells (FC) deliver an initial discharge capacity of 106~mAh~g$^{-1}$ at an average voltage of 3.2~V and retained $\sim$79\% capacity after 100 cycles at 2~C, highlighting their potential for high-energy sodium-ion storage. Post-cycling analyses confirms that the cathode preserves its structural and morphological integrity even after 1000 cycles at 10~C, signifying remarkable long-term stability.

\section{\noindent Experimental}

{\noindent {\bf Materials synthesis:}} The NVP-HE powder sample was synthesized using a sol-gel method followed by thermal treatment. All chemical reagents used were of analytical grade (Sigma-Aldrich) and employed without further purification. Initially, 9.5 mmol of NH$_4$VO$_3$ and citric acid (constituting 80\% of the total precursor weight) were dissolved in 60 mL of deionized water (DI) and heated to 70\degree C under continuous stirring until a blue, transparent solution was obtained. Subsequently, 0.1 mmol each of Ni(NO$_3$)$_2\cdot$6H$_2$O, C$_{20}$H$_{28}$O$_8$Zr, Cr(NO$_3$)$_2\cdot$9H$_2$O, Al(NO$_3$)$_3\cdot$9H$_2$O, and (NH$_4$)$_6$Mo$_7$O$_{24}\cdot$4H$_2$O were sequentially introduced to the above solution under continuous stirring. Following this, 15 mmol of NH$_4$H$_2$PO$_4$ and 15 mmol of NaNO$_3$ were added to the solution. To enhance electrical conductivity, carbon nanotubes (CNTs, 5 wt.\% of the total precursor weight) were ultrasonically dispersed in 20 mL of DI for 40 mins and then incorporated into the reaction mixture. The solution was heated at 80\degree C until a viscous gel was formed and then vacuum dried overnight at 120\degree C in an oven. The dried powder is calcined at 350\degree C for 4 hrs and 800\degree C for 8 hrs in a tubular furnace under Ar and 5\% H$_2$ environment. The reference NVP was prepared with the same procedure without inclusion of HE dopants. 

{\noindent {\bf Physical characterization:}} The X-ray powder diffraction (XRD) measurements were carried out over a 2$\theta$ range of 10--80$^\circ$ using a PANalytical X'Pert$^3$ powder diffractometer equipped with Cu K$_\alpha$ radiation ($\lambda$ = 1.5406~\AA), operating at 45 kV and 40 mA. The Rietveld refinement of XRD patterens was performed using the FullProf software to verify phase purity and to extract crystallographic parameters. Sodium-ion migration behavior was examined using the \textit{SoftBV GUI tool} based on the bond valence site energy (BVSE) method \cite{Wong_CM_21} where the input data for the simulation was sourced from the Rietveld-refined crystallographic file. The SoftBV calculates energy landscapes by evaluating bond valence mismatches, and the resulting isosurfaces representing favorable sodium-ion pathways were visualized using \textit{VESTA} software \cite{Momma_JAC_11}. The Raman spectroscopy was performed using a Renishaw inVia confocal Raman microscope equipped with a 532~nm laser, a 2400 lines/mm grating, and a laser power of 10 mW to investigate the vibrational modes of the samples. Surface morphology was studied via scanning electron microscopy (SEM, TESCAN MAGNA), while elemental composition and spatial distribution were examined using EDS coupled with a JEOL JSM-7800F Prime microscope equipped with Oxford ED. Further, the ICP-MS measurements were performed for elemental analysis using an Agilent 7900 instrument equipped with Ultra High Matrix Introduction (UHMI) and operated in helium (He) mode to minimize polyatomic interferences, after digesting the powdered samples in freshly prepared aqua regia. The HR-TEM and selected area electron diffraction (SAED) analyses were carried out using a JEM-ARM200F NEOARM microscope operated at an accelerating voltage of 200~keV to explore fine microstructural characteristics at high resolution. The X-ray photoemission spectroscopy (XPS) was performed using a Kratos AXIS Supra instrument equipped with a monochromatic Al K$\alpha$ X-ray source (photon energy = 1486.6 eV) to analyze the oxidation states of the constituent elements. The core-level spectra are calibrated using the C~1$s$ reference peak positioned at 284.6~eV. The inelastic background signal is subtracted using the Tougaard method, and the core-level spectra  are fitted using Voigt function (a convolution of Gaussian and Lorentzian profiles) with a shape factor of 0.6, utilizing \textit{IgorPro} software. 

{\noindent {\bf CR~2032-type cell fabrication:}} The cathode slurry was prepared by homogeneously dispersing 70~wt.\% active material, 20~wt.\% Super~P conductive carbon, and 10~wt.\% polyvinylidene difluoride (PVDF) binder in N-methyl-2-pyrrolidone (NMP) solvent. The obtained slurry was uniformly cast onto battery grade aluminum foil and subsequently dried under vacuum at 80\degree C for 12~hrs. The resulting electrodes were calendared for density optimization and punched into 12~mm discs, yielding mass loading of 2--3~mg\,cm$^{-2}$. The CR~2032-type cells were assembled in an argon-filled glovebox (UniLab Pro SP, MBraun), with oxygen and moisture levels maintained below 0.01~ppm where sodium metal served as the counter/reference electrode, and glass fiber (GB100-R, Advantec) was utilized as a separator. The electrolyte comprised 1~M NaPF$_6$ dissolved in propylene carbonate (PC) with 2~wt.\% fluoroethylene carbonate (FEC) as an additive. Commercial hard carbon (HC, particle size $\sim$9~$\mu$m) was procured from Shaldong Gelon LIB Co. Ltd. and used as the anode material for full-cell fabrication. The anode slurry was prepared with a weight ratio of 8:1:1 (HC:PVDF:Super~P) and cast onto copper foil to form the working electrode. 

{\noindent {\bf Electrochemical measurements:}} Galvanostatic charge--discharge (GCD) measurements were conducted using a Neware battery testing system over a voltage window of 2.0--4.3~V vs.\ Na/Na$^+$. The electrochemical characterizations, including CV, GITT, and EIS, were performed using a Biologic VMP-3 workstation. The CV was performed within 2.0--4.3~V at scan rates ranging from 0.05 to 1.0~mV\,s$^{-1}$. The GITT employed current pulses of 10~minutes (0.1~C) followed by relaxation intervals of 60~minutes within the voltage range of 2.0--4.3~V vs.\ Na/Na$^+$. The EIS measurements were carried out at open-circuit potential using a 10~mV {\it a.c.} perturbation within the frequency range of 0.01~Hz to 10$^5$~Hz. The viability of EIS data was assessed via Kramers-Kronig consistency tests using \textit{Lin-KK tool} software \cite{Schoenleber_EA_14}, ensuring residuals of real and imaginary components remained within $\sim$1\%. The DRT analysis of EIS spectra was performed using a MATLAB-based open-source tool (\textit{DRT-tools}) \cite{Song_PRL_18, Plank_JPS_24} where Tikhonov regularization coupled with a non-linear least squares fitting approach was employed, with the regularization parameter fixed at 0.0001 and a second-order derivative constraint. The radial basis function’s full width at half maximum (FWHM) was set to 0.5 for optimized spectral fitting. Subsequently, the DRT peaks were fitted using Gaussian non-linear fitting in \textit{Igor Pro} software to extract relaxation time and polarization resistance. 

\begin{figure*}[htbp]
\includegraphics[width=7.3in]{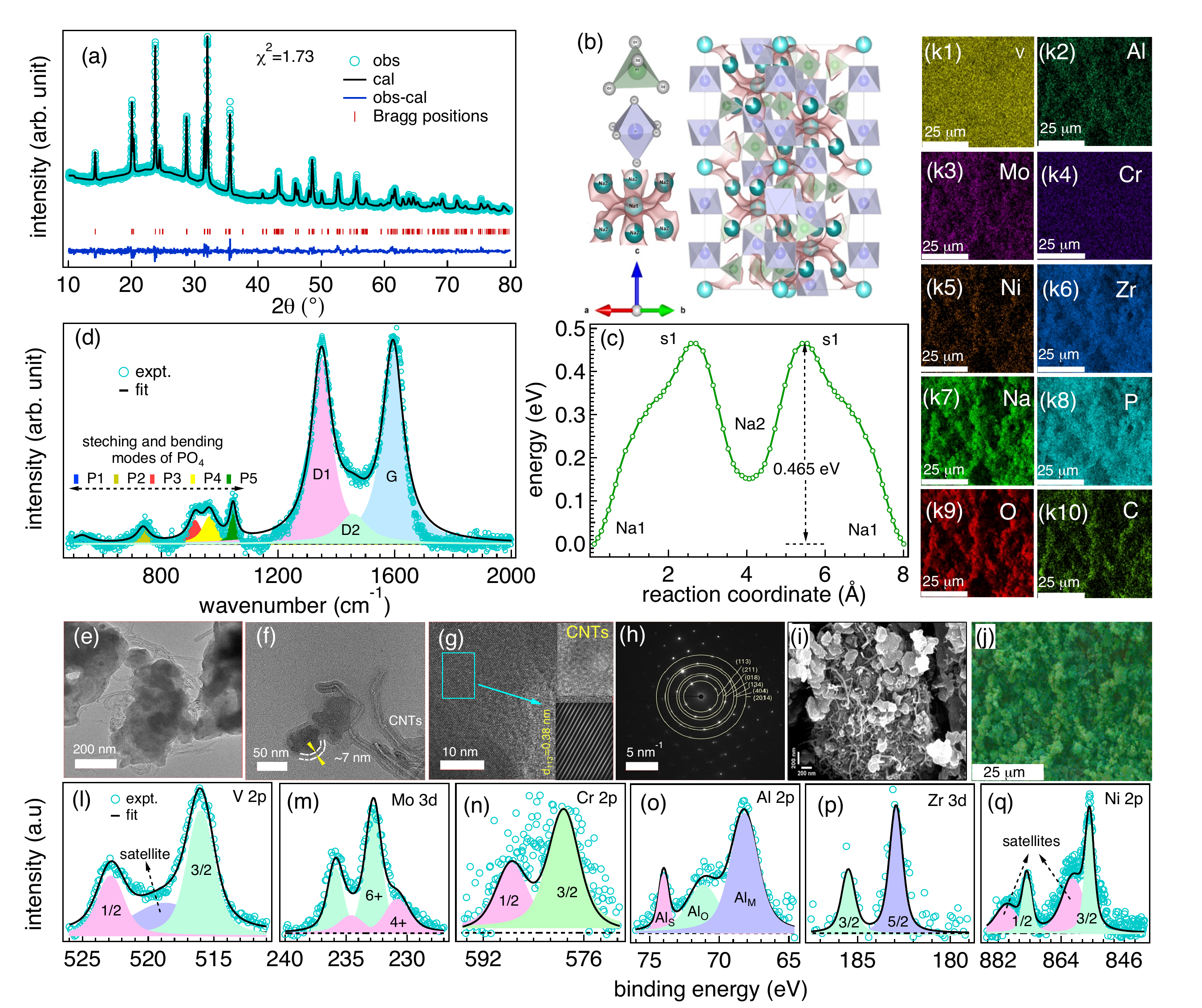}
\caption {(a) The XRD pattern with Rietveld refinement profile of the NVP-HE sample; (b) an isosurface representation of Na$^+$ ion diffusion pathways in the NVP-HE structure, and (c) the corresponding energy profile of migration barriers along the minimum energy path Na1--Na2--Na1, calculated using the BVSE approach; (d) The Raman spectrum; (e-g) the HR-TEM images, (h) the SAED pattern; (i) the FESEM image, (j) overlay image of elemental mapping, and (k1-k10) individual elemental mapping of constituent elements of NVP-HE sample; the core-level XPS spectra of the (l) V 2$p$, (m) Mo 3$d$, (n) Cr 2$p$, (o) Al 2$p$, (p) Zr 3$d$, and (q) Ni 2$p$.}
\label{F1}
\end{figure*}

\section{\noindent Results and discussion} 

The Rietveld refinement of the XRD pattern shown in Fig.~\ref{F1}(a) confirms rhombohedral (R$\bar{3}$c) structure with lattice parameters $a=b=$ 8.7369 ${\rm \AA}$, $c=$ 21.8378 ${\rm \AA}$, and the interaxial angles of $\alpha$ = $\beta$ = 90\degree, $\gamma$ = 120\degree, resulting in a unit cell volume of 1443.6 ${\rm \AA}^3$. The crystallographic details, provided in Tables S1 and S2 of \cite{SI}, are closely matching with previous reports \cite{Park_Nat.Mat_2025, Kang_CC_2024, Wang_JMCA_2024}. The NVP-HE framework, displayed in Fig.~\ref{F1}(b), is constructed from corner-sharing VO$_6$ octahedra and PO$_4$ tetrahedra forming the characteristic [V$_2$(PO$_4$)$_3$]$^{3-}$ lantern unit, which accommodates two partially occupied sodium sites where Na(1), located at the 6$b$ position and coordinated by six oxygen atoms, contributes to structural stability, while Na(2), situated at the 18$e$ position and coordinated by eight oxygen atoms, facilitates Na$^+$ transport within the framework. Notably, the (012) reflection of NVP-HE exhibits a shift of $\sim$0.18\degree{} toward lower 2$\theta$ values compared to pristine NVP (Fig.~S1(a,b) of \cite{SI}), indicating a slight lattice expansion, as well as a marginal elongation of the V--O bond from 2.037~\AA{} to 2.040~\AA{}, accompanied by a contraction of the P--O bond from 1.518~\AA{} to 1.509~\AA{}. The elongated V--O bond signifies mild distortion of the VO$_6$ octahedra, which is expected to locally widen Na$^+$ diffusion bottlenecks and reduce migration energy barriers. Also, the shortened P--O bond indicates reinforcement of the PO$_4$ polyanionic framework, enhancing structural stability during repeated Na$^+$ insertion and extraction \cite{Sun_ACS_nano_25, Kang_CC_2024, Su_AEM_2025}. Furthermore, bond-valence site energy (BVSE) analysis identifies energetically favorable 3D Na$^+$ migration pathways, as illustrated in Fig.~\ref{F1}(b), where Na(1) and Na(2) sites are interconnected through Na(1)--Na(2)--Na(1) or Na(2)--Na(1)--Na(2) sequences [left panel of Fig.\ref{F1}(b)], forming a continuous and percolating network that facilitates long-range ion diffusion. The estimated activation energy for Na$^+$ migration in NVP-HE is found to be approximately 0.465 eV, as illustrated in Fig.~\ref{F1}(c), marginally lower than the 0.483 eV observed for the NVP [see Fig.~S1(c)], indicating the broader diffusion channels \cite{Park_Nat.Mat_2025}. The deconvoluted Raman spectrum, displayed in Fig.~\ref{F1}(d), identify the PO$_{4}$$^{3-}$ vibrational modes at $\sim$528.2 cm$^{-1}$ and $\sim$737.8 cm$^{-1}$, which can be attributed to symmetric and asymmetric bending vibrations, while the modes around 915.5 cm$^{-1}$, 965.8 cm$^{-1}$, and 1046.2~cm$^{-1}$ correspond to symmetric and asymmetric stretching vibrations, confirming the integrity of the polyanionic framework \cite{Singh_PRB_2024}. Moreover, the observed D$1$ band at 1349.0~cm$^{-1}$ arises from disorder-induced vibrations, indicating structural defects within the carbon coating. The D$2$ band at 1457.8~cm$^{-1}$ is attributed to the presence of amorphous carbon and the G band at 1594.4~cm$^{-1}$ corresponds to the in-plane stretching vibrations of sp$^2$-bonded carbon atoms in graphitic domains \cite{Shen_AMI_2016}. The ratio (I$_{D1}$/I$_{G}$ = 0.88) reflects a moderate degree of disorder in the carbon structure \cite{Zhang_AMI_2019}. Similar spectral features are observed in the Raman spectrum of NVP, although with comparatively lower disorder in the graphitic carbon, as shown in Fig.~S2 of \cite{SI}.

\begin{figure*}[htbp]
\includegraphics[width=7.1in]{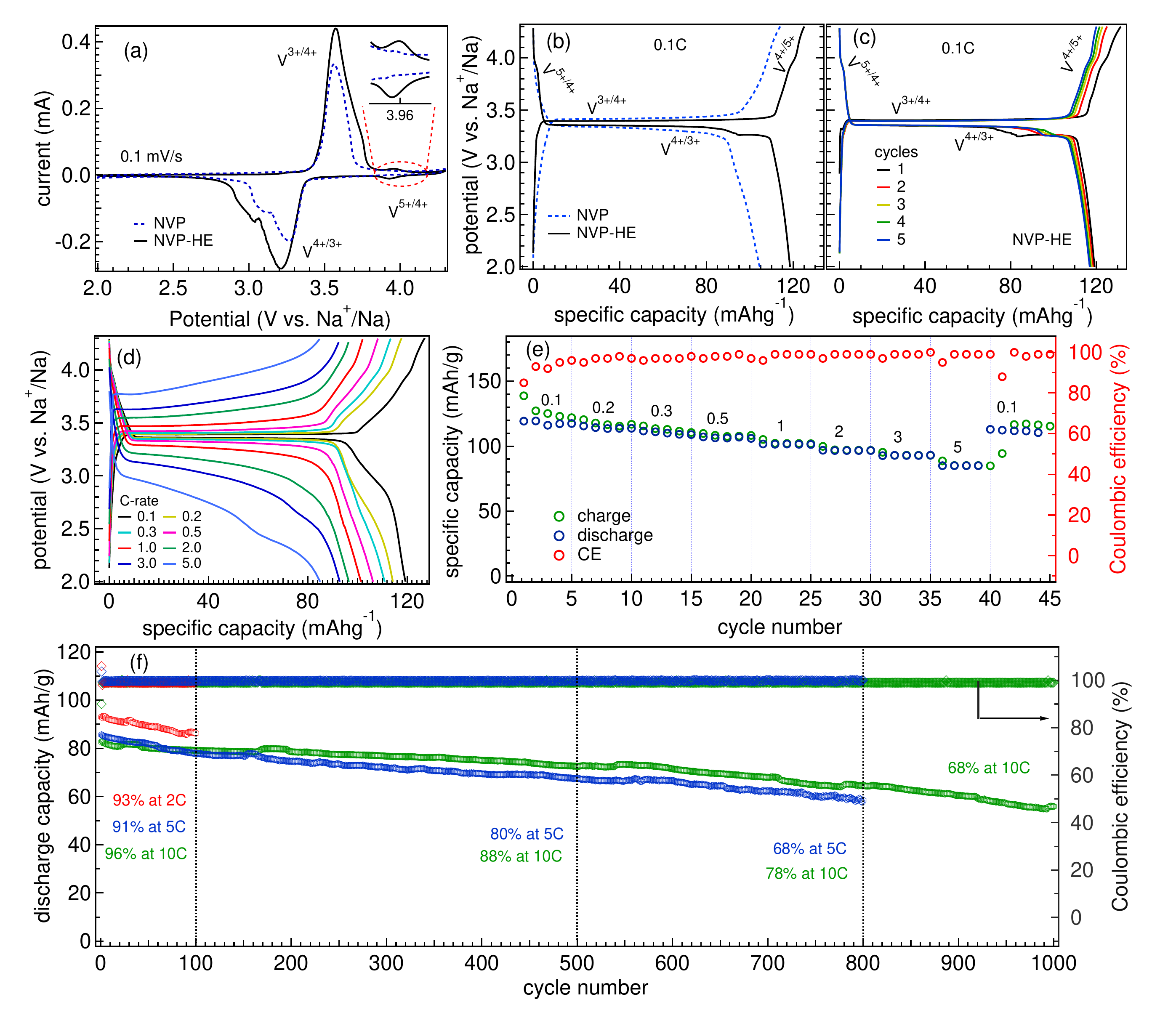}
\caption {(a) The CV curves at 0.1 mV s$^{-1}$, (b) the GCD profiles at 0.1~C for the second cycle of NVP and NVP-HE in 2.0--4.3~V; (c) the first five cycles of GCD profile of NVP-HE at 0.1C; (d) the rate performances; (e) the corresponding GCD profiles for second cycle at each current rate; (f) the cycling performances of NVP-HE at the current rates of 2~C, 5~C, and 10~C.}
\label{F3_a}
\end{figure*}

Further, the high-resolution transmission electron microscopy (HR-TEM) analysis of NVP-HE powder sample reveals the microstructures with non-uniform particle shapes, as shown in Figs.~\ref{F1}(e, f). The images confirm a uniform carbon coating of $\sim$~7 nm thickness on the surface of the cathode particles, along with the successful incorporation of CNTs, as evident from Figs.~\ref{F1}(e, f). The lattice-resolved HR-TEM images in Fig.~\ref{F1}(g) show an inter-planar spacing of 0.38~nm corresponding to the (113) plane of the rhombohedral phase, while the top inset displays CNT lattice fringes, confirming the successful integration of the CNT matrix into the composite system. The distinct diffraction spots observed in SAED, as shown in Fig.~\ref{F1}(h), can be assigned to the (113), (211), (018), (134), (404), and (2014) planes, corroborating the XRD findings discussed earlier and confirming the crystallinity of the NVP-HE phase. Fig.~\ref{F1}(i) display the  FE-SEM images of NVP-HE sample, exhibiting irregularly shaped particles ranging from $\sim$~100~nm to a few microns. The composite elemental overlay is presented in Fig.~\ref{F1}(j), while the individual elemental distributions are shown in Figs.~\ref{F1}(k1--k10). The corresponding maps demonstrate uniform and homogeneous distribution of different elements across the entire sample. The quantitative elemental composition obtained from the EDS measurements is provided in Fig.~S3 of \cite{SI}, validating the successful multi-elemental doping within the lattice framework. Also, the ICP-MS analysis confirms the presence of all intended dopants (see Table S3 of \cite{SI}). The Na/V ratio remains closely aligned with the targeted stoichiometry, suggesting that the dopants predominantly substitute for V$^{3+}$ sites within the lattice framework. The V 2$p$ core-level in Fig.~\ref{F1}(l) shows the spin-orbit splitting peaks at 516.0~eV and 522.9~eV confirming V$^{3+}$ oxidation state along with a satellite feature at 518.7~eV~\cite{Zhang_AMI_2019, Li_AS_22, Liao_ESM_25}. The Mo~3$d_{5/2}$ peaks appear at 230.8~eV and 232.8~eV for the for Mo$^{4+}$ and Mo$^{6+}$, and the corresponding Mo~3$d_{3/2}$ peaks are located at 234.4~eV and 235.8~eV, respectively \cite{Liao_ESM_25, Madhav_Small_2025} reveal a mixed-valence state. The Cr~2$p$ core-level in Fig.~\ref{F1}(n) displays peaks at 579.2~eV and 587.6~eV, consistent with the Cr$^{3+}$ oxidation state~\cite{Li_AS_22, Gu_AM_2022, Liao_ESM_25}. In Fig.~\ref{F1}(o), the Al~2$p$ spectrum exhibits three distinct peaks centered at 68.2~eV (Al$_{M}$), 71.2~eV (Al$_{O}$), and 74.0~eV (Al$_{S}$), which can be assigned to metallic Al, Al$^{3+}$, and surface-bound component, respectively~\cite{Gu_AM_2022, Liao_ESM_25}. The Zr~3$d$ core-level in Fig.~\ref{F1}(p) shows peaks at 182.9~eV and 185.5~eV, attributed to Zr~3$d_{5/2}$ and Zr~3$d_{3/2}$, respectively, confirming the Zr$^{4+}$ oxidation state \cite{Liao_ESM_25}. The Ni~2$p$ spectrum in  Fig.~\ref{F1}(q) displays peaks at 855.9~eV (Ni~2$p_{3/2}$) and 873.8~eV (Ni~2$p_{1/2}$), along with their respective satellite peaks, which are indicative of the Ni$^{2+}$ valence state~\cite{Liao_ESM_25}. The XPS spectra for the C 1$s$, Na 1$s$, O 1$s$, and P 2$p$ are presented in Figs.~S1(h-k) of \cite{SI}.

Now we examine the electrochemical performance of the NVP-HE cathode and first compare the CV curves, recorded at a scan rate of 0.1~mV~s$^{-1}$ within 2.0--4.3~V. Both the NVP and NVP-HE cathodes exhibit redox peaks corresponding to the V$^{3+}$/V$^{4+}$ couple at 3.56/3.27~V and 3.57/3.21~V, respectively. In successive cycles, the NVP-HE electrode displays a smaller potential gap and higher peak overlap, indicating superior redox reversibility compared to the NVP (see Figs.~S2(a, b) of \cite{SI}). Interestingly, an additional redox pair at 3.96/3.93~V is observed in NVP-HE, which confirms the activation of V$^{4+}$/V$^{5+}$ couple via high-entropy doping. The GCD profiles recorded at 0.1~C, see Figs.~\ref{F3_a}(b, c) and Fig.~S2(c) of \cite{SI}, demonstrate a flat voltage plateau associated with the V$^{3+}$/V$^{4+}$ redox couple, at around 3.41/3.34~V. In addition, the NVP-HE exhibits a distinct plateau near 3.95~V, attributed to the V$^{4+}$/V$^{5+}$ redox couple  \cite{Wang_JMCA_24, Kang_CC_2024}, consistent with CV observations. Notably, the observed polarization in NVP-HE ($\sim$50~mV) is found to be significantly lower than that of pristine NVP ($\sim$90~mV), as shown in Fig.~\ref{F3_a}(b), indicating enhanced charge-transport properties and accelerated electrochemical kinetics \cite{Hou_ESM_24, Ren_Nano_En_2016, Zhou_Nano_En_24, Zhang_ESM_24}. As a result, the NVP-HE cathode delivers a higher specific discharge capacity of 119~mAh~g$^{-1}$ at 0.1~C as compared to 105~mAh~g$^{-1}$ for the NVP, benefitting from the contribution of the activated V$^{4+}$/V$^{5+}$ redox reaction. Fig.~\ref{F3_a}(c) presents the GCD curves of NVP-HE at 0.1~C for the first five cycles, which show nearly identical profiles with a stabilized discharge capacity of $\sim$117~mAh~g$^{-1}$. Moreover, the relatively low Coulombic efficiency observed during the first cycle can be attributed to electrolyte decomposition at high voltage and formation of the solid electrolyte interphase (SEI) layer \cite{Ponrouch_EES_2012}. Further, the NVP-HE cathode exhibits 85~mAh~g$^{-1}$ at 5~C with Coulombic efficiencies approaching 100\%, and the capacity recovery of 95\% upon returning back to 0.1~C, see Figs.~\ref{F3_a}(d, e). The corresponding GCD profiles, shown in Fig.~\ref{F3_a}(d), demonstrate increasing voltage polarization with higher C rate, which is attributed to kinetic limitations such as sluggish Na$^+$ diffusion and elevated charge-transfer resistance \cite{Jiang_JMCA_2016, Gu_AM_2022}. The long cycling test of NVP-HE at 10~C showed 68\% retention after 1000 cycles, as displayed in Fig.~\ref{F3_a}(f). We find that the capacity retention improves at 10 C (78\% after 800 cycles) as compared to 5 C (68\%), which can be attributed to the reduced time of the electrode at high potentials during faster cycling, thereby limiting side reactions \cite{Grundish_CM_2020, Pati_JMCA_2022}. 

\begin{figure*}[htbp]
\includegraphics[width=7.1in]{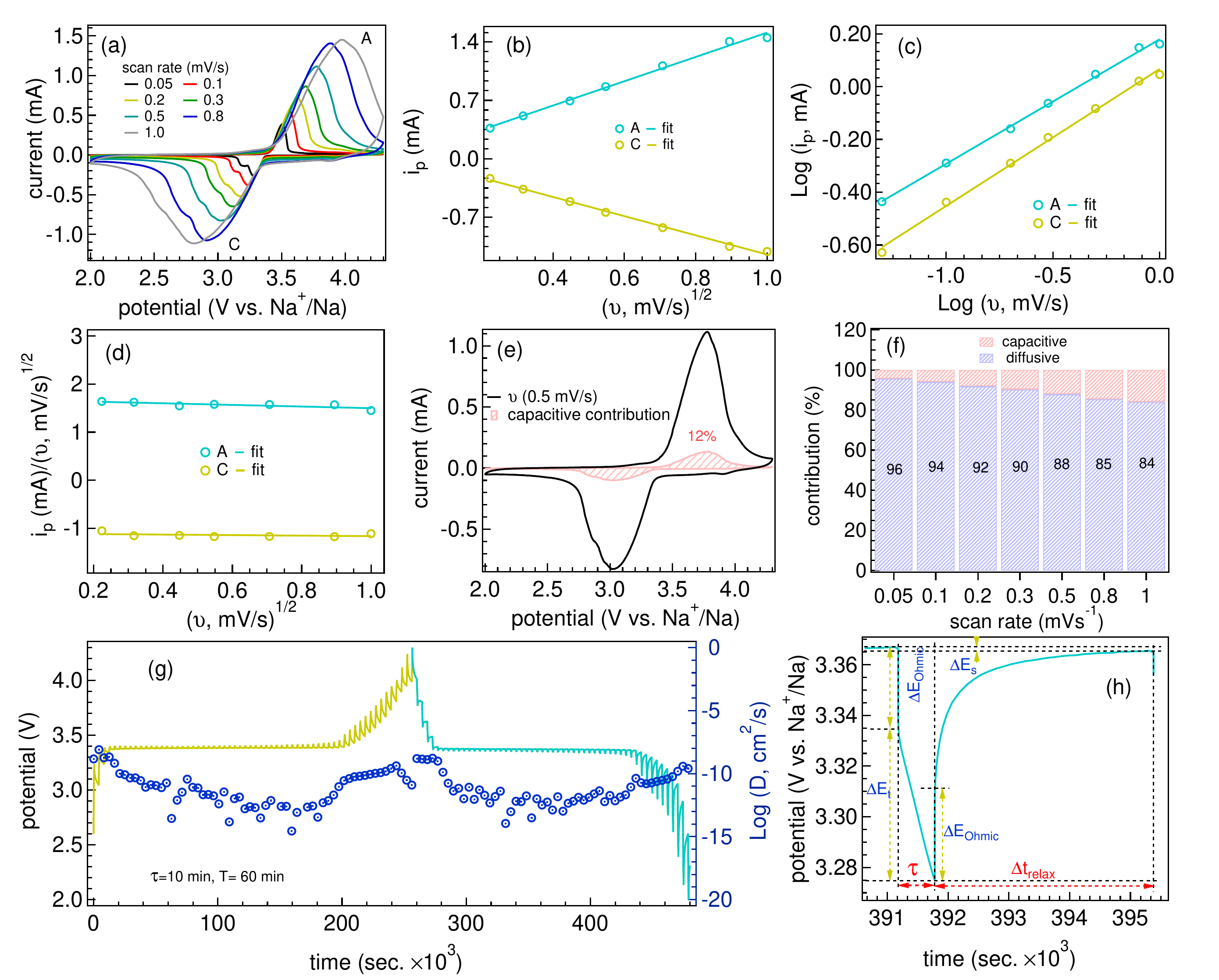}
\caption {(a) The CV curves of NVP-HE at various scan rates 0.05--1.0 mV s$^{-1}$; (b) a linear relationship between peak current (i$_p$) and the square root of the scan rate ($\upsilon$$^{1/2}$); (c) a linear fitting of log(i$_p$) versus log($\upsilon$), and (d) i$_p$/$\upsilon$$^{1/2}$ versus $\upsilon$$^{1/2}$ for anodic and cathodic peaks; (e) the capacitive contribution at 0.5 mV/s, highlighted by the shaded area; (f) the distribution of capacitive and diffusion-controlled current components at various scan rates; (g) the GITT profile and diffusion coefficient recorded with a pulse time of 10 mins and rest time of 60 mins at 0.1~C in 2.0--4.3 V range after formation cycles; (h) depiction of various parameters during a galvanostatic titration step involving a 10 mins current pulse followed by a 60 mins rest.} 
\label{F3_b}
\end{figure*}

Further, the CV curves presented in Fig.~\ref{F3_b}(a) show a shift in anodic peaks toward higher potentials and cathodic peaks toward lower potentials, which indicate an enhancement in the polarization with increasing the scan rate. The rise and broadening in redox peak intensity suggests a growing capacitive contribution linked to electric double-layer effects \cite{Schoetz_EA_2022, Gogotsi_ACSN_18}. The V$^{4+}$/V$^{5+}$ redox couple displays suppressed activity at higher scan rates, likely due to intrinsic kinetic limitations and restricted electronic conductivity \cite{Schoetz_EA_2022, Zhou_CS_2024}. Note that the Faradaic processes are governed by diffusion-controlled intercalation and redox reactions within the electrode bulk, whereas non-faradaic contributions arise from surface-driven phenomena such as pseudo-capacitance and electric double-layer formation \cite{Gogotsi_ACSN_18}. To further quantify the diffusion-controlled contributions, diffusion coefficient for anodic and cathodic peaks is estimated using the Randles–\v{S}ev\v{c}ik equation \cite{Su_AEM_2025}.
\begin{equation}
i_p = (2.69 \times 10^5) A D^{\frac{1}{2}} C\eta^{\frac{3}{2}} \upsilon^{\frac{1}{2}}
\label{RS}
\end{equation}
where $i_p$ (mA) denotes the peak current, $A$ (cm$^2$) is the electrode surface area, $D$ (cm$^2$/s) represents the diffusion coefficient, $C$ (mol/cm$^3$) is the bulk concentration of Na$^+$ ions, $\eta$ is the number of electrons involved in the redox process, and $\upsilon$ (mV~s$^{-1}$) refers to the scan rate. The peak current is extracted from the CV curves for both anodic and cathodic scans (A and C) and plotted against the square root of the scan rate in Fig.~\ref{F3_b}(b). Using equation~\ref{RS} and the slopes obtained from linear fitting, the diffusion coefficient for the anodic and cathodic peaks is found to be $6.0 \times 10^{-12}$ and $3.7 \times 10^{-11}$~cm$^2$/s, respectively. It is important to note that the Randles--\v{S}ev\v{c}ik (RS) model considers Faradaic contributions and exclude the pseudo-capacitive effects. However, the broadening of redox peaks observed in the CV curves indicates a hybrid charge storage mechanism involving both Faradaic and pseudo-capacitive processes \cite{Su_AEM_2025}. To further evaluate the kinetic contributions at different scan rates, a $\log(i)$ versus $\log(\upsilon)$ plot is constructed and linearly fitted using the power-law in Fig.~\ref{F3_b}(c)~\cite{Schoetz_EA_2022}.
\begin{equation}
i = a\,\upsilon^b \quad \text{or} \quad \log(i) = \log(a) + b\,\log(\upsilon)
\label{power_law}
\end{equation}
where $i$ denotes the current, $\upsilon$ represents the scan rate, and $a$ and $b$ are empirical fitting parameters. The value of $b$ reflects the charge storage mechanism, with $b=$ 0.5 indicates a diffusion-controlled process governed by the Randles–\v{S}ev\v{c}ik (RS) model, $b=$ 1 signifying a surface-controlled  capacitive behavior, and intermediate values denote a combination of capacitive and diffusion-driven contributions, characteristic of finite-length diffusion \cite{Schoetz_EA_2022, Gogotsi_ACSN_18}. The calculated $b$ values of 0.47 (anodic) and 0.52 (cathodic) imply a diffusion-dominated charge storage mechanism in the present care \cite{Ding_ACS_Nano_24, Zhang_AMI_2019, Ding_JPS_24}. To quantitatively separate the contributions from surface and diffusion controlled processes in the total current response, the peak current ($i_p$) can be expressed as a combination of both mechanisms, following the methodology proposed by Liu \textit{et al.} \cite{Liu_JES_98} and later adopted by Choi \textit{et al.} \cite{Choi_NatRM_20}; 
\begin{equation}
\frac{i}{\upsilon^{\frac{1}{2}}} = k_1 \upsilon^{\frac{1}{2}} + k_2
\end{equation}
In this model, $k_1$ and $k_2$ represent the capacitive and diffusion-controlled contributions, respectively, where the values are extracted from the linear fitting of $\frac{i}{\upsilon^{1/2}}$ versus $\upsilon^{1/2}$, as shown in Fig.~\ref{F3_b}(d). The calculated contributions are displayed in Fig.~\ref{F3_b}(f) at various scan rates, and Fig.~\ref{F3_b}(e) displays both the contribution at a scan rate of 0.5 mV~s$^{-1}$, with only $\sim$~12\% capacitive nature. These results indicate that the charge storage mechanism is predominantly governed by diffusion-controlled kinetics over the entire range of studied scan rates \cite{Park_CEJ_2021}. 

Note that the GITT analysis is widely regarded as a more precise method for determining the diffusion coefficient, as it is conducted close to the equilibrium conditions, thereby reducing the influence of kinetic and capacitive distortions. Prior to perform the GITT measurements, the fresh cell was tested up to five formation cycles of charge and discharge at 0.1~C to ensure electrode stability. The time-dependent GITT response is illustrated in Fig.~\ref{F3_b}(g). The testing protocol involved initiating the cell charge to the upper voltage limit of 4.3~V by applying a current pulse for 10 mins duration ($\tau$), followed by a relaxation interval of 60 mins. This rest period allow the sodium ions to redistribute uniformly within the electrode and enabled the cell voltage to approach a quasi-equilibrium open-circuit voltage (OCV), characterized by an almost zero voltage change rate (dE/dt~$\rightarrow$~0~V/s). This sequence of charge pulse and relaxation was subsequently repeated during discharge until the voltage reached the lower cutoff at 2.0~V. Fig.~\ref{F3_b}(h) depicts the voltage profile for one discharge titration step indicating all the parameters involved. To determine the diffusion coefficient at thermodynamic equilibrium, Fick’s second law can be reduced under specific assumptions, including the adoption of a semi-infinite diffusion case \cite{Nickol_JES_20}:
\begin{equation}
D_{\text{Na}^+} = \frac{4}{\pi \tau} \left( \frac{m_{B}V_{M}}{M_{B}A} \right)^2 \left(\frac{\Delta E_S}{\tau \left( \frac{dE_{\tau}}{d\sqrt{\tau}} \right)} \right)^2; \tau \ll \frac{L^2}{D_{{Na}^+}}
\label{FICK1}
\end{equation}
here, $m_B$ and $M_B$ are the weight (in grams) and the molar mass (in g/mol) of the cathode active material. $V_M$ is the molar volume (cm$^3$/mol), $L$ is the electrode thickness ($\mu$m), and $A$ denotes the electrode’s active surface area (cm$^2$). The symbol $\tau$ signifies the time duration of the applied current pulse (s), $\Delta E_S$ represents the difference in potential prior to and following the pulse, and $\Delta E_\tau$ is the voltage gap between the equilibrium and maximum potentials at the end of the current pulse, as indicated in Fig.~\ref{F3_b}(h). When the transient voltage varies linearly with ${\tau}^{1/2}$, this equation simplifies to:
\begin{equation}
D_{\text{Na}^+} = \frac{4}{\pi \tau} \left( \frac{m_{B}V_M}{M_{B}A} \right)^2 \left( \frac{\Delta E_S}{\Delta E_{\tau}} \right)^2
\label{FICK2}
\end{equation} 
The average apparent sodium-ion diffusion coefficients, $D_{\text{Na}^+}$, are estimated to be 7.5$\times$$10^{-11}$~cm$^2$~s$^{-1}$ during charging and  3.1$\times$$10^{-11}$~cm$^2$~s$^{-1}$ during discharging, see Fig.~\ref{F3_b}(g). These values suggest that Na$^+$ ions diffuse through the cathode material at similar rates in both directions within this voltage window, whereas an order of magnitude difference is observed in RS-derived $D_{\text{Na}^+}$ values. This can be attributed to the fact that RS analysis includes pseudo-capacitive contributions inherently embedded in peak currents, reducing its specificity to diffusion processes \cite{Pati_JPS_2024}.

\begin{figure*}[htbp]
\includegraphics[width=7.1in]{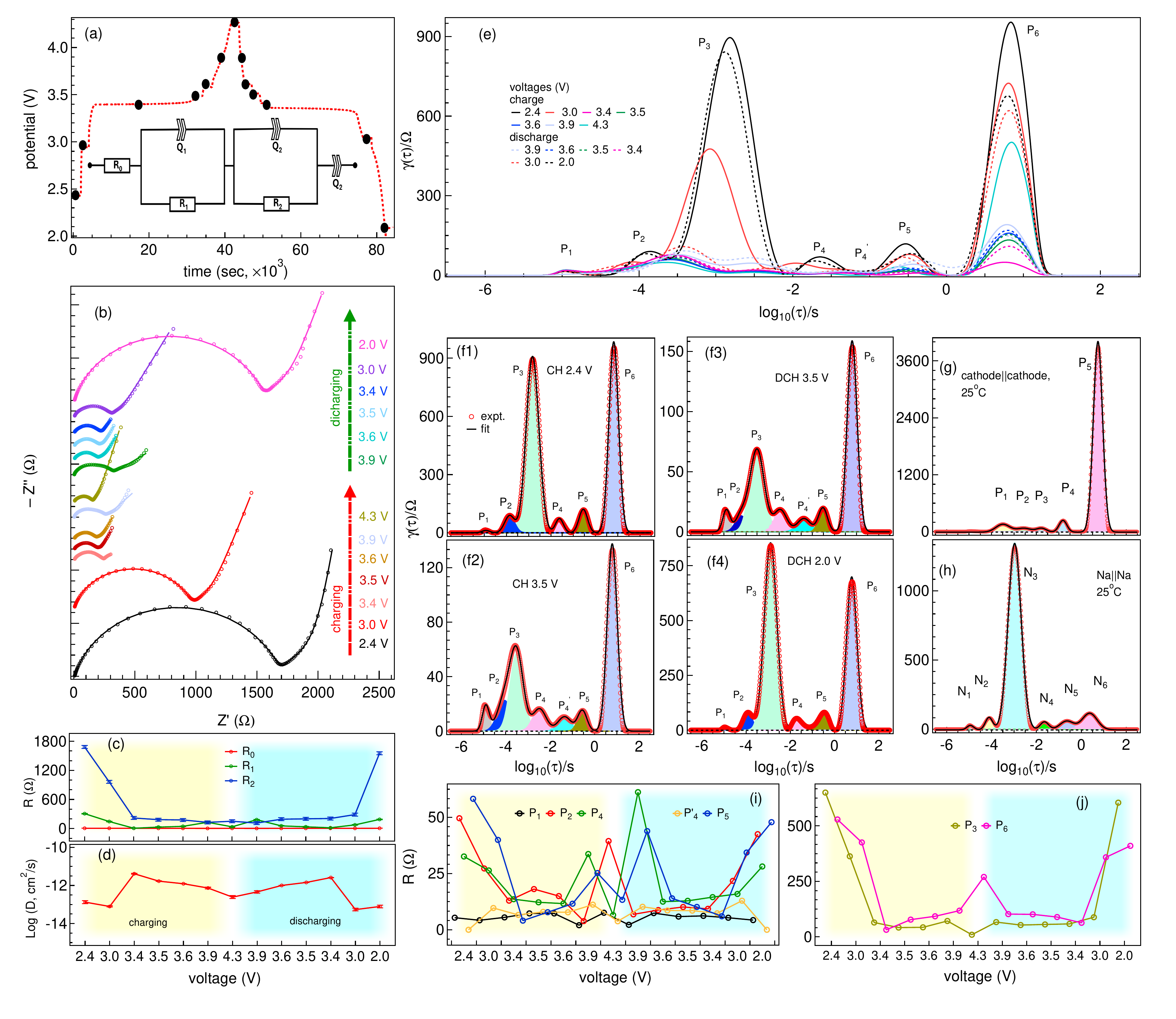}
\caption {(a) The GCD profile of NVP-HE cathode with points indicating where the EIS measurements are performed; the corresponding $in-situ$ EIS spectra in (b) recorded at various potential values during charging and discharging; (c) the resistance as a function of voltage obtained from the EIS fitting and (e) sodium-ion diffusion coefficient values extracted from the Warburg region of the corresponding EIS spectra; (e) the simulated DRT profiles derived from EIS measured at different voltages; (f1--f4) the fitted DRT profiles for selected voltages, the DRT profiles of symmetric cells of NVP-HE$\parallel$NVP-HE in (g) and Na$\parallel$Na in (h), (i, j) the resistance values extracted from peak deconvolution of DRT profiles, shown as a function of voltage.} 
\label{F3_c}
\end{figure*}

Moreover, it is equally important to probe the electrode-electrolyte interfacial behavior to gain deeper understanding of the underlying reaction mechanisms. Therefore, the EIS measurements are performed at selected voltages representing different stages of the sodiation and de-sodiation process, as depicted in Fig.~\ref{F3_c}(a), with the corresponding Nyquist plots shown in Figs.~\ref{F3_c}(b). Prior to the EIS measurements, the cell was subjected to five formation cycles at 0.1~C to ensure a stable electrode-electrolyte interface. In order to minimize voltage drift during open-circuit conditions, the cell was allowed only a short rest period of 10 s before recording each EIS spectrum, with drift current compensation enabled. The EIS spectra in Figs.~\ref{F3_c}(b) exhibit a depressed semicircle extending over the high- and mid-frequency regions, which can be deconvoluted into two overlapping semicircles, followed by a straight line in the low-frequency domain. As illustrated by the equivalent circuit shown in the inset of Fig.~\ref{F3_c}(a), the initial intercept on the $x-$axis corresponds to the Ohmic resistance of the cell (R$_0$), which represents the combination of contacts, electrolyte, and separator. The first semicircle in the Nyquist plot corresponds to the solid electrolyte interphase (SEI), modeled by R$_1$ and Q$_1$, and originates from the resistive and capacitive response of the interfacial layer formed by electrolyte decomposition products at the electrode surface, where Q denotes the constant phase element (CPE), which accounts for non-ideal capacitive behavior. The second semicircle reflects the charge-transfer process at the electrode--electrolyte boundary, modeled by R$_{2}$ and Q$_2$, which is strongly influenced by electrochemical reaction kinetics and is sensitive to factors such as Na site occupancy, surface chemistry, structural transitions, electronic structure, and particle morphology \cite{Plank_JPS_24, Zhang_AFM_2026, Ding_JPS_24}. At lower frequencies, the inclined line is assigned to the Warburg region (Q$_3$), which accounts for the solid-state diffusion of Na-ions within the cathode matrix \cite{Li_NatCommun_2019}. The variation of the fitted resistance parameters (see Table~S6) as a function of cell potential is shown in Fig.~\ref{F3_c}(c), extracted using the equivalent circuit model. The solution resistance (R$_{0}$) remains relatively stable over the entire voltage range, indicating that the electrolyte conductivity does not undergo any pronounced change during the charge--discharge process. In contrast, the charge-transfer resistance (R$_2$) exhibits the strongest voltage dependence and thus dominates the overall resistance, whereas the interfacial resistance (R$_1$) shows a comparatively weaker response. During the charging process, as the potential increases from 2.4~V to 4.3~V, a pronounced decrease is observed in both R$_{1}$ (from $\sim307~\Omega$ to $\sim30~\Omega$) and R$_{2}$ (from $\sim1680~\Omega$ to $\sim184~\Omega$). This reduction reflects the gradual removal of Na$^{+}$ from the Na$_{2}$ site of the NVP-HE framework, consistent with the activation of the V$^{3+}$/V$^{4+}$ redox reaction and the enhancement of Na$^{+}$ transport kinetics within the electrode \cite{Wu_AdvMater_2025}. Upon subsequent discharge,  as the voltage decreases to 2.0~V, both R$_1$ and R$_2$ increase significantly (R$_{1} \approx189~\Omega$, R$_2 \approx1554~\Omega$), indicating reoccupation of the Na$_2$ sites, accumulation of interfacial resistance and sluggish ion transport as it approaches deep sodiation \cite{Wu_AdvMater_2025}. In order to estimate the intrinsic diffusion behavior of Na$^{+}$ within the electrode bulk at different electrode potentials, it is necessary to examine the low-frequency response in the Warburg region of the Nyquist plots,  
using the following equations~\cite{Sapra_ESM_2025}:
\begin{equation}
D_{Na^+} = \frac{R^2 T^2}{2 A^2 \eta^4 F^4 C^2 \sigma^2}
\label{diff}
\end{equation}
\begin{equation}
Z' = R_s + R_{ct} + \sigma \omega^{-0.5}
\label{war}
\end{equation}
where \(R\) is the gas constant, \(T\) the temperature, \(A\) the electrode’s geometric area, \(\eta\) the number of sodium ions involved, \(F\) the Faraday constant, \(C\) the sodium concentration (mol/cm\(^3\)), and \(\sigma\) the Warburg coefficient. The Warburg coefficient ($\sigma$) is determined using equation~\ref{war}, where $Z'$ corresponds to the real impedance and $R_{s}$ and $R_{ct}$ represent the solution and charge transfer resistances, respectively, and its value can be obtained from the slope of the linear fit of $Z'$ versus $\omega^{-1/2}$ (rad~s$^{-1}$)$^{-1/2}$, as shown in Fig.~S2(d) of \cite{SI}. The calculated Na$^{+}$ diffusion coefficient ($D_{\text{Na}^+}$) fall in the range of $10^{-11}$ to $10^{-12}$~cm$^{2}$~s$^{-1}$ in the redox-active regions, as displayed in Fig.~\ref{F3_c}(d), which found to be in good agreement with GITT measurements discussed above.

Here, it is important to realize that equivalent circuit fitting provides quantitative information but often oversimplifies complex spectra by reducing them to a few lumped elements, masking overlapping processes. To address this limitation and achieve greater physical resolution, the impedance spectra are analyzed using the DRT method \cite{Hahn_Batteries_2019}. The DRT does not require a predefined model, instead, it reconstructs the response by treating the system as an infinite network of resistor--capacitor elements with varying time constants using infinite Voigt circuit, produces a distribution function that links resistance contributions to their characteristic relaxation times, thereby distinguishing processes such as charge transfer, interface polarization, and ion diffusion \cite{Hahn_Batteries_2019}. Since this inversion is mathematically ill-posed and noise-sensitive, Tikhonov regularization was applied with an optimized parameter of 0.0001, providing a balance between resolution and stability. Each peak in the resulting spectra reflects a specific electrochemical phenomenon, with its position denoting the characteristic relaxation time and its area corresponds to the associated polarization resistance. We use high-quality impedance data, acquired over a broad frequency range (100~mHz to 100~kHz) for reliable DRT analysis. Prior to DRT processing, the validity of the EIS data is confirmed using the Kramers-Kronig (K-K) transformation, which tests causality and consistency \cite{Schoenleber_EA_14}. The K-K residuals remain well below 1\% across, in Fig.~S3(b),  the scanned frequencies, verifying the physical integrity of the data, which ensures that subsequent DRT results accurately reflect real system dynamics. The DRT curves obtained at various charging (solid lines) and discharging (dotted lines) voltages, as presented in Fig.~\ref{F3_c}(e), exhibit well-resolved peaks across the examined time constant domain, which provides mechanistic insights into processes that remain convoluted in conventional Nyquist plots \cite{Lu_Joule_2022}. The assignment of the DRT peaks for the NVP-HE half-cell is systematically done based on characteristic time scales \cite{Hahn_Batteries_2019, Lu_Joule_2022, Semerukhin_E_Acta_2024, Zhao_JPS_2022, Doyle_JES_1993} through direct comparison with the DRT profiles of symmetric cells, namely NVP-HE$\parallel$NVP-HE and Na$\parallel$Na measured at 25\degree C, see Figs.~\ref{F3_c}(g, h). 

The fitted DRT profiles, displayed in Figs.~\ref{F3_c}(f1--f4) at selected voltages and for more details see Figs.~S3(a1--a9) and Table~S7 of \cite{SI}, provide precise time constants and associated resistance values; for example, peaks P$_1$ and P$_2$ ($\tau \approx 12~\mu$s and 0.1~ms at OCV) are assigned to particle-particle and particle-current collector contact resistances within the NVP-HE electrode and CNT network. These peaks remain nearly voltage-independent, see Fig.~\ref{F3_c}(e) and Table~S7, indicating non-faradaic contact processes \cite{Hahn_Batteries_2019}. The P$_3$ peak, centered at $\tau \approx 1.5$~ms (at OCV), exhibits a pronounced voltage dependence characteristic of charge-transfer kinetics, governed by the interfacial over-potential and local Na$^+$ activity \cite{Plank_JPS_24}. Accordingly, the P$_3$ peak is assigned to the charge-transfer resistance at the Na metal--electrolyte interface \cite{Zhao_JPS_2022}, which is consistent with the DRT response of Na--Na symmetric cell (N$_3$ peak), see Fig.~\ref{F3_c}(h). The P$_4$ peak located at $\tau \approx22$~ms (at OCV) in Fig.~\ref{F3_c}(f1)  is assigned to Na$^+$ migration through the SEI  \cite{Hahn_Batteries_2019}. The P$_5$ peak observed at $\approx$0.3~s is attributed to charge-transport processes at the cathode-electrolyte interface  \cite{Semerukhin_E_Acta_2024, Zhao_JPS_2022}. The P$_6$ peak at 6.6~s is associated with solid-state Na$^+$ diffusion within the cathode structure, consistent with the DRT profile of the NVP-HE cathode symmetric cell, see Fig.~\ref{F3_c}(g). This confirms that the migration of sodium ions through the bulk of the electrode is a relatively slower phenomenon as compared to interfacial electrochemical processes \cite{Semerukhin_E_Acta_2024}. 

We observe significant change in the DRT peak intensity, which indicate the polarization driven electrochemical response. During the charging, the resistive contributions associated with most relaxation processes gradually decrease with increasing potential, reflecting progressive activation of interfacial reactions, as displayed in Figs.~\ref{F3_c}(i, j). Notably, a slight scattering is observed in the values of resistances around 3.9--4.3 V, which can be ascribed to the intrinsically sluggish V$^{4+}$/V$^{5+}$ redox transition [see  Fig.~\ref{F3_c}(i, j) and Table~S7]. Further, the low-frequency diffusion-associated process exhibits a slight intensification at 4.3~V, indicative of emerging mass-transport constraints under deep desodiation \cite{Doyle_JES_1993}. During the discharge, resistance values increases with the voltage when reaches to around 2.0 V, as depicted in Fig.~\ref{F3_c}(i, j). We also observe a new peak ($P_4^{\prime}$), which is visible in Figs.~\ref{F3_c}(f2, f3) and Figs.~S3 of \cite{SI}), suggesting a metastable interfacial layer at intermediate voltages. Overall, the DRT analysis identifies charge transfer and solid-state diffusion as the primary kinetic bottlenecks,  that signifies Na$^+$ intercalation is most efficient during the central redox plateau, but becomes significantly impeded by elevated energetic barriers as the system approaches sodiation/desodiation limits, as illustrated in Figs.~\ref{F3_c}(i, j). Here, the resistance and relaxation times at equal voltages across charge-discharge indicate high kinetic reversibility, confirming the electrochemical robustness and structural stability of the NVP-HE framework. 

\begin{figure*}
	\includegraphics[width=7.2in]{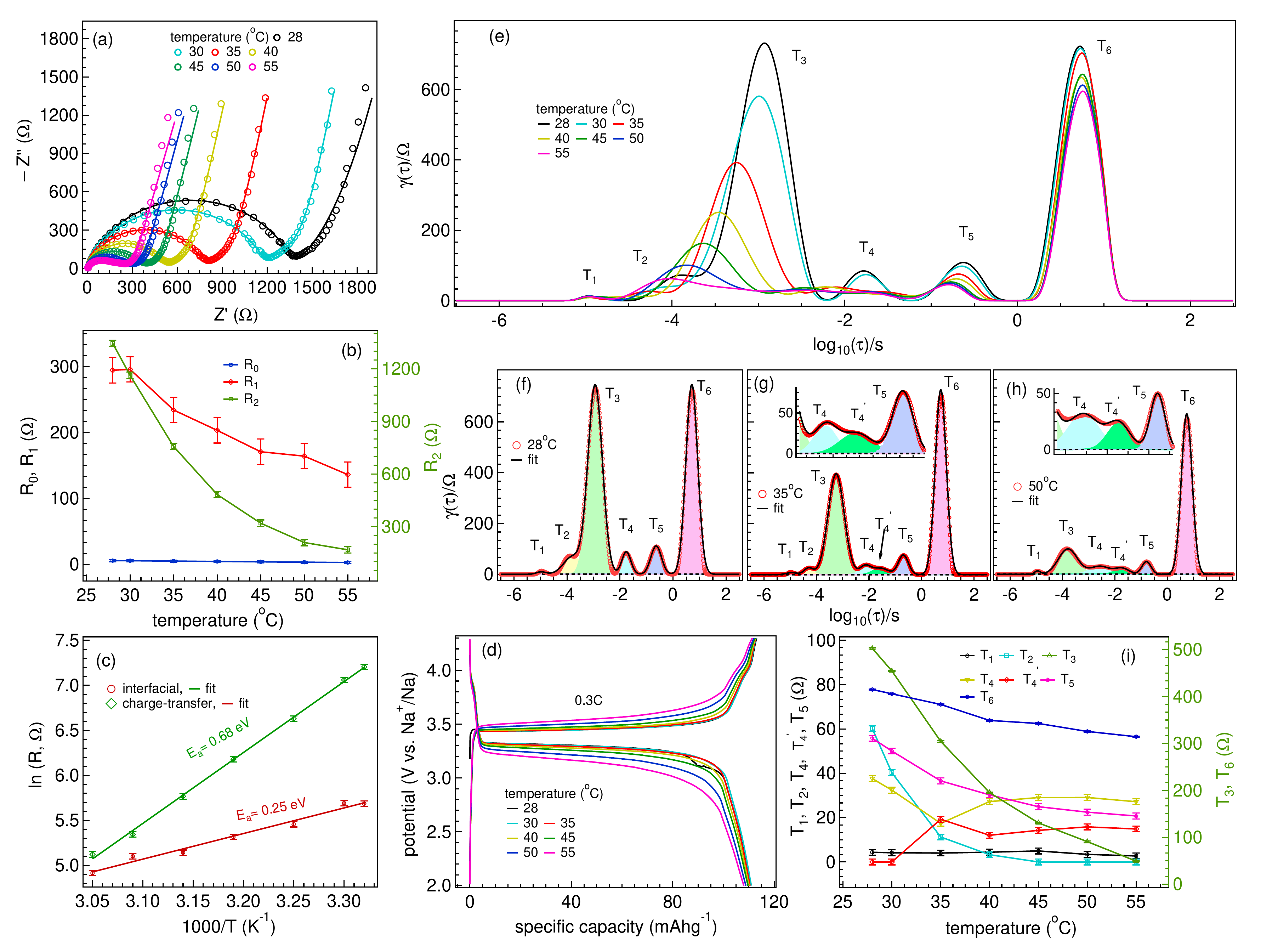}
	\caption {(a) The EIS spectra of NVP-HE cathode recorded at different temperatures from 28\degree C to 55\degree C, (b) the resistance values derived from equivalent circuit fitting of the EIS spectra, (c) the Arrhenius plot corresponds to interfacial and charge-transfer resistances, (d) the GCD profile of NVP-HE cathode at various temperatures measured at 0.3~C; (e) the simulated DRT profiles extracted from each EIS spectrum; (f--h) the fitted DRT profiles at 28\degree C, 35\degree C, and 50\degree C, respectively; (i) the resistance values extracted from peak deconvolution of DRT profiles corresponding to individual peaks at different temperatures.} 
	\label{F3_d}
\end{figure*} 

Moreover, the EIS measurements are performed at various temperatures in the range of 28--55\degree C and at the open-circuit voltage (OCV) after formation cycles, see Figs.~\ref{F3_d}(a). The data are fitted using the same equivalent circuit shown in the inset of Fig.~\ref{F3_c}(a), and the extracted resistance values are illustrated in Fig.~\ref{F3_d}(b) and Table~S8. The solution/contact resistance (R$_0$) and SEI resistance (R$_1$) decrease moderately; for example, R$_0$ from 5.7~$\Omega$ to 2.9~$\Omega$ and R$_1$ from $\sim$295~$\Omega$ to 136~$\Omega$ in the temperature range of 28--55\degree C. However, the faradaic charge-transfer resistance (R$_2$) changes significantly, i.e., from $\sim$1345~$\Omega$ to 168~$\Omega$. This reduction in resistance values at elevated temperatures can be attributed to enhanced ionic conductivity, reduced electrolyte viscosity, and the formation of more conductive interfacial layers, which enhance ionic transport and charge--transfer kinetics, thereby improving overall electrode performance \cite{Zhou_JPS_2019}. Further, higher thermal energy accelerates charge--transfer mecanism at the electrode--electrolyte interface by lowering activation barriers, as described by the Arrhenius relation, while simultaneously promoting faster interfacial reactions, collectively minimizing resistive losses \cite{Zhou_JPS_2019, Illig_JES_2012}. The Arrhenius equation is  given by \( R = A \exp\left(E_{\mathrm{a}} / k_{\mathrm{B}} T\right) \), where, A is pre-exponetial factor and E$_a$ is the activation energy. The extracted activation energy values reveal a clear kinetic separation between the resistance components. The SEI resistance exhibits a relatively low activation energy (E$_a$) of 0.25~eV, indicating facile Na${^+}$ migration through the interphase layer, whereas the charge-transfer resistance is associated with a substantially higher E$_a$ value of 0.68~eV, see Fig.~\ref{F3_c}(c). This elevated barrier signifies that Na${^+}$ insertion into the NVP-HE framework is intrinsically activated and constitutes the rate-limiting step of the electrochemical process \cite{Scanlan_EleActa_25, Busche_NatChem_16}. 

To assess the practical impact of temperature on electrochemical performance, the GCD profiles are measured for the NVP-HE cathode at 0.3~C in 2.0--4.3~V range, as displayed in Fig.~\ref{F3_d}(d). The electrode delivers a fairly stable discharge capacity of $\sim$110~mAh~g$^{-1}$ between 28\degree C and  55\degree C. However, the polarization increases progressively above 35\degree C, exhibiting a contrasting trend to the EIS results, which suggest lower resistance/polarization and consequently, improved kinetics at elevated temperatures. To elucidate this inconsistency and deconvolute the underlying electrochemical processes, temperature-dependent DRT analysis is performed, as presented in Fig.~\ref{F3_d}(e). Prior to the DRT processing, the reliability of the EIS data is verified by Kramers–Kronig transformation in the frequency range of 100~mHz to 100~kHz, which yielded residuals below 1.3\% [see Fig.~S4(i)], ensuring that the subsequent DRT results accurately reflect the true system dynamics \cite{Schoenleber_EA_14}. The fitted profiles are shown in Figs.~\ref{F3_d}(f--h) and the corresponding resistance values are displayed in Fig.~\ref{F3_d}(i) and summarized in Table~S9. The values of charge-transfer resistance at anode-electrolyte interface (peak T$_3$ at $\tau \sim$1~ms) exhibit significant decrease from 502~$\Omega$ to 49~$\Omega$ when temperature increases from 28\degree C to 55\degree C, indicating a significant enhancement of interfacial charge-transfer kinetics. All other resistances like cathodic solid-state diffusion (T$_6$ at $\sim$5.1~s), T$_4$ ($\tau \sim$17~ms, charge transport at the anode--electrolyte interface), T$_5$ ($\tau \sim$0.24~s, charge transport at the NVP-HE cathode--electrolyte interface) exhibit slow decrease in resistance values with temperature. Notably, a new relaxation peak, denoted as T$_4^{\prime}$, marked in Figs.~\ref{F3_d}(g, h), emerges at 35\degree C having the resistance values of $\sim$15--19~$\Omega$ upto 55\degree C. The appearance of T$_4^{\prime}$ coincides with the increased polarization in GCD profiles in Fig.~\ref{F3_d}(d) suggesting the formation of a thermally induced secondary interphase at $\ge$35\degree C \cite{Ould_ChemComm_2025, Ravdel_JPS_2003}. 

\begin{figure}[h]
	\includegraphics[width=3.5in]{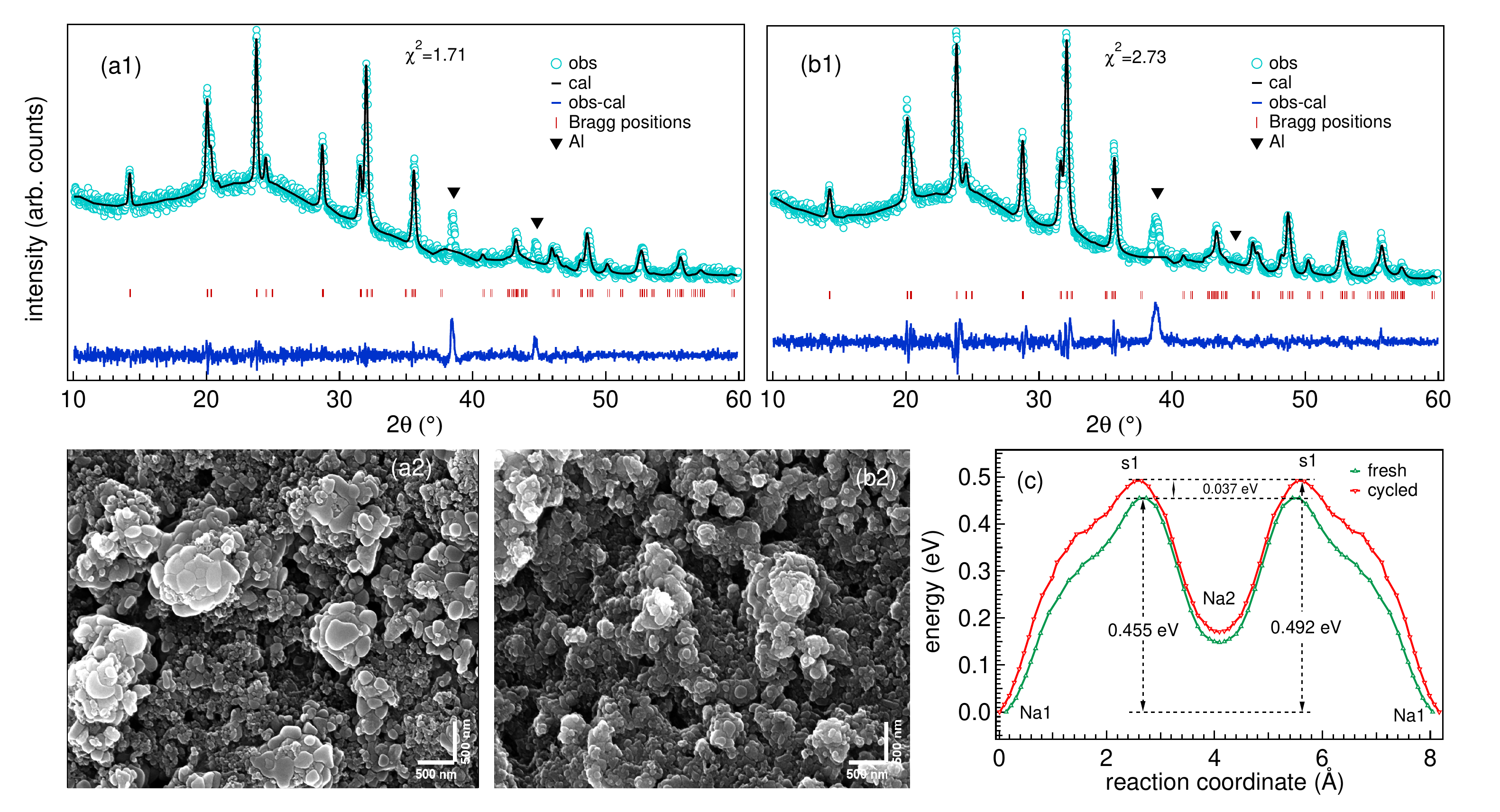}
	\caption {The Rietveld refinement of XRD patterns of NVP-HE: (a1) a fresh electrode with corresponding (a2) FE-SEM image; (b1) an electrode after 1000~cycles at 10~C, with corresponding (b2) FE-SEM image; (c) the calculated migration energy profiles, using the BVSE approach, along the Na1-Na2-Na1 diffusion pathways. }
	\label{F3_g}
\end{figure} 

To investigate the structural and morphological evolution of the NVP-HE cathode after long cycling (1000 cycles at 10~C), the post-mortem XRD and FE-SEM measurements are carried out. The cycled half-cells were disassembled, and the recovered cathode material thoroughly rinsed with dimethyl carbonate to remove residual electrolyte species, followed by drying inside the glovebox. The Rietveld refinement of XRD patterns of the pristine and cycled electrodes, with goodness-of-fit values of 1.7\% and 2.7\%, respectively, are shown in Figs.~\ref{F3_g}(a1, b1), which remain consistent with the active phase (R$\bar{3}$c). The detailed refinement parameters are summarized in Tables~S4 and S5. A noticeable diffraction peaks originating from the Al current collector are also observed in both cases, marked by downward triangles in the respective profiles. Notably, the crystal structure of NVP-HE remains essentially intact, with negligible lattice distortion/peak shifting even after 1000 cycles at 10~C, confirming excellent structural stability. The Na$^{+}$ migration activation energies estimated from BVSE analysis are 0.455~eV (fresh) and 0.492~eV (cycled), as presented in Fig.~\ref{F3_g}(c), where the slight increase can be attributed to lattice contraction ($\Delta$V = 9.52~${\rm \AA}^3$), leading to compressive strain, narrower diffusion channels, and consequently slightly higher migration barriers \cite{Wu_AdvMater_2025, Wang_JMCA_2024}. Furthermore, the FE-SEM images of pristine and cycled electrodes, as displayed in Figs.~\ref{F3_g}(a2, b2) demonstrate that there is no notable particle aggregation/cracking, suggesting that the NVP-HE particles effectively accommodate lattice volume fluctuations during repeated intercalation/deintercalation, thereby suppressing structural degradation \cite{Zhang_AFM_2024}. These findings highlight the synergistic pillar effect of high-entropy elements in NVP to stabilize the structure and morphology by mitigating phase transitions and lattice distortions during cycling \cite{Garcia_ESM_24, Ding_ACS_Nano_24}.

\begin{figure}[h]
	\includegraphics[width=3.4in]{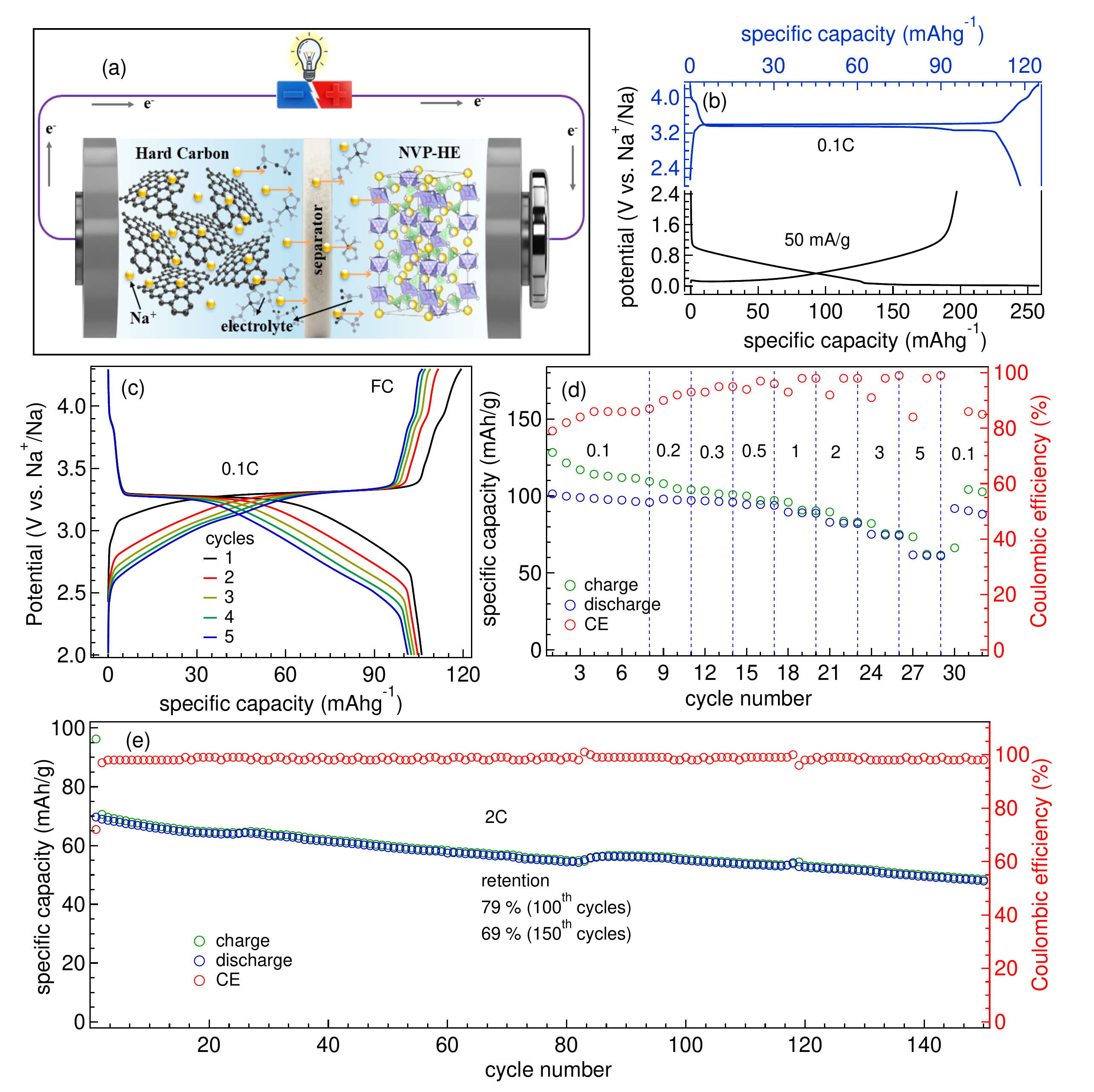}
	\caption {(a) Schematic illustration of the operating principle of the HC$\parallel$NVP-HE full cell, (b) GCD profile of hard carbon anode at current rate of 50~mA/g in voltage window of 0.01 to 2.5~V and of NVP-HE cathode at 0.1~C in the voltage window of 2.0 to 4.3~V; (c) GCD profile of HC$\parallel$NVP-HE full cell at 0.1~C in the voltage window of 2.0--4.3~V, (d) rate performance and (e) cycling performance at 2~C of full cell.}
	\label{FC}
\end{figure}

Finally, to evaluate the practical applicability of the NVP-HE cathode, full cells (FC) are assembled using a hard carbon (HC) anode, as illustrated in Fig.~\ref{FC}(a). The mass balance in NVP-HE$\parallel$HC full cells is maintained at $\sim$1.2 to mitigate irreversible capacity degradation \cite{Sapra_AMI_2024}. Prior to the full-cell fabrication, the HC was electrochemically pre-sodiated in half cell configuration at 50~mA~g$^{-1}$ for three cycles within 0.1--2.5~V range to compensate for irreversible Na$^{+}$ loss, after which the cells were disassembled to retrieve the sodiated HC \cite{Sapra_JMCA_25}. The GCD profiles of individual NVP-HE and HC half cells are shown in Fig.~\ref{FC}(b), delivering initial discharge capacities of 119~mAh~g$^{-1}$ at 0.1~C and 258~mAh~g$^{-1}$ at 50~mA~g$^{-1}$, respectively. The corresponding GCD curve of the full cell in 2.0--4.3~V range measured at 0.1~C is presented in Fig.~\ref{FC}(c) up to five cycles. The initial discharge capacity is 106~mAh~g$^{-1}$, which decreases slightly to 101~mAh~g$^{-1}$ after five cycles, accompanied by an initial Coulombic efficiency of 89\%. The relatively low Coulombic efficiency is attributed to the limited sodium inventory and irreversible Na$^{+}$ consumption during SEI/CEI formation \cite{Sapra_AMI_2024}. Notably, the full cell exhibited favorable rate capability, delivering specific capacities of 101, 98, 97, 94, 89, 83, 75, and 62~mAh~g$^{-1}$ at current rates of 0.1, 0.2, 0.3, 0.5, 1, 2, 3, and 5~C, respectively, with 91\% retention when comes back to 0.1 C rate, see Fig.~\ref{FC}(d)). The corresponding cycling performance is shown in Fig.~\ref{FC}(e) exhibiting $\sim$79\% capacity retention after 100 cycles and $\sim$69\% capacity retention after 150 cycles at 2~C. Furthermore, the average operating voltage of the full cell is found to be $\sim$3.2~V at 0.1~C, yielding an energy density of 326~Wh~kg$^{-1}$ (based on cathode mass), underscoring the promise of NVP-HE for high-energy-density SIBs.  

\section{\noindent Conclusions}

In summary, a high-entropy doped cathode Na$_3$V$_{1.9}$(CrMoAlZrNi)$_{0.1}$(PO$_4$)$_3$ was successfully synthesized via a sol--gel method, yielding a phase-pure material with homogeneous elemental distribution. Comprehensive structural and microstructural analyses confirmed the structural integrity, particle morphology, and local bonding environments. The incorporation of five mixed-valence dopants in trace amounts, verified by EDS, ICP-MS, and XPS, enabled high-entropy mixing at the vanadium site, modulating the lattice, enhancing electronic and ionic conductivity, and activating V$^{4+}$/V$^{5+}$ redox couple. This resulted a reversible capacity of 119~mAh~g$^{-1}$ at 0.1~C, and minimal polarization (0.05~V) with excellent rate capability up to 5~C and remarkable cycling stability with 68\% at 10~C after 1000 cycles. The sodium-ion diffusion coefficients found to be consistent, determined via CV, EIS, and GITT in the range of \(10^{-11}\)–\(10^{-13}~\mathrm{cm^2\,s^{-1}}\). The {\it in-situ} EIS coupled with DRT analysis, which deconvolutes various underlying polarization mechanisms of the electrochemical system, provided deeper insights into charge-transfer and transport processes and explained the increased polarization observed at elevated temperatures in the GCD profile of NVP-HE half cells. Further, the full cells with HC anode delivered an energy density of 326~Wh~kg$^{-1}$ at an average voltage of $\sim$3.2~V based on cathodic mass and retained $\sim$79\% capacity after 100 cycles at 2~C. The post-mortem XRD and FE-SEM analyses confirmed the retention of structural integrity and morphology even after 1000 cycles at 10 C, highlighting the NVP-HE cathode as a promising candidate for high-energy, long-life SIBs.  

\section{\noindent Acknowledgments}

MKS gratefully acknowledges the MoE for fellowship support through IIT Delhi and Dr. Simranjot K. Sapra for valuable discussions. RSD acknowledges the Department of Science and Technology (DST) for financial support for the sodium-ion battery project through the “DST–IIT Delhi Energy Storage Platform on Batteries” (Project No. DST/TMD/MECSP/2K17/07), as well as the Science and Engineering Research Board (SERB, now ANRF) through a Core Research Grant (File No. CRG/2020/003436). The authors also acknowledge IIT Delhi for providing access to XPS, FE-SEM, and HR-TEM facilities at the Central Research Facility (CRF), along with XRD and Raman spectroscopy facilities at the Department of Physics. 



\end{document}